\documentclass[12pt]{article}
\usepackage{amssymb,amsmath}
\begin{document}

\renewcommand{\thefootnote}{\fnsymbol{footnote}}

\begin{center}
{\large {\bf Simplified Version of the Nucleon and Delta-Isobar
Description \\[0pt] }}
 \end{center}

\begin{center}
\textbf{V. D. Tsukanov}
\end{center}
\begin{center}{\it Institute of Theoretical Physics, National Science Center}\\[0pt]
{\it ''Kharkov Institute of Physics and Technology''}\\[0pt]
{\it 61108, Kharkov, Ukraine}\\[0pt]
\end{center}

\begin{quotation}
{\small {\rm Considering synchronous spin-isospin rotation as a
collective motion a simplified scheme of the nucleon and delta
isobar description is formulated.}}
\end{quotation}

\renewcommand{\thefootnote}{\arabic{footnote}} \setcounter{footnote}0
\vspace{1cm}

\section{Introduction}

Spin properties of the baryons within the frame of the Skyrme
model are studied since early eighties \cite{ANW}. The key moment
of these researches is the use of the moduli space approximation
(MSA). Such an approach is not a consecutive theory. It is of
artificial character and assumes obligatory existence of the
static soliton solution. Degeneration parameters of these
solutions are regarded as the collective variables while studying
the internal dynamics of solitons. Such a choice of the collective
variables does not make possible to describe various nucleon
resonances. Besides, it does not allow to take into account the
dynamic deformation of solitons. Solution of the mentioned
problems can be found only on the base of complete dynamic
description (CDD) \cite{Ts}. This description assumes the use of
the variational approach to the collective variables formalism and
the employment of the complete set of cyclic coordinates as such
variables. Thus, while describing nucleon and its resonances, the
configurational space consists of group parameters describing spin
rotation and group parameters related to isotopic rotation. This
theory shows that at the classical level the nucleon and
delta-isobar states are exact solutions of the motion equations at
which rotations in the spin and isospin space are synchronous. The
last circumstance allows to formulate a simplified theory, the
configurational space of which is described by one set of SU(2)
group parameters, reflecting a synchronous rotation of particles
in isotopic and spin spaces. Obviously, this theory cannot
describe the entire variety of nucleon resonances, but it will
allow to solve some of the problems mentioned above as applied to
the nucleon and delta-isobar states. The creation of this scheme
along with the assessment of its possibilities is given below.

\section{Dynamics of the synchronous spin-isospin rotation}

Lagrangian of the SU(2) invariant model of nucleon, written in
terms of group parameters $\phi ^{i}(x,t)$ can be presented in the
form

\begin{equation}
L(\phi ,\dot{\phi})=1/2<g_{ik}(\phi )\dot{\phi}^{i}\dot{\phi}^{k}>-H(\phi
,0)\,,\qquad <A>\equiv \int d^{3}x\ A(x)\,.  \label{a}
\end{equation}

In order to provide the possibility for moderate soliton rotation,
the Lagrangian includes mass term and depending on concrete form
of the kinetic matrix $g_{ik}(\phi )$ and the potential term
$H(\phi ,0)$ can describe both SU(2) sigma-model and corresponding
Skyrme model. The Lagrangian (\ref a) are invariant with respect
to isotopic $\phi(x)\rightarrow T\phi (x)$ and space rotations
$\phi (x)\rightarrow \phi (Tx)$, where $T$ is three-dimensional
orthogonal matrix of the adjoint representation of SU(2). The
conserved dynamic functionals of the isotopic and ordinary spin
associated with the invariance properties mentioned have the form
\begin{equation}
\begin{split}
&\boldsymbol{I}=-i<\hat{\boldsymbol{I}}\phi
^{i}g_{ik}\dot{\phi}^{k}>,\qquad
\hat{I}_{ks}^{i}=-i\varepsilon _{iks}\ , \\
&\boldsymbol{J} =-i<\hat{\boldsymbol{l}}\phi
^{i}g_{ik}\dot{\phi}^{k}>,\qquad \hat{l}_{ks}^{i}(x)=-i\delta
_{ks}(\boldsymbol{x}\times \partial /\partial
\boldsymbol{x})_{\,i}\,, \label{b}
\end {split}
\end{equation}
where $\hat{\boldsymbol{I}}$, $\hat{\boldsymbol{l}}$ are the
kinematic operators of the isotopic spin and the angular momentum.
The purpose of this work is to create a simplified version of the
nucleon and delta-isobar description as compared to the CDD scheme
of the work \cite{Ts}. Therefore, let us use the results of this
work and connect the collective motion only with the states that
correspond to synchronous rotation of the system in isotopic and
spin spaces. Then the corresponding collective variables can be
introduced by means of the substitution
\begin{equation}
\phi (x)=T\varphi (Tx)\,,  \label{d}
\end{equation}
where the role of these variables is played by group parameters
$a^{i}$ of the matrix $T$. Unlike the work \cite{Ts} in which spin
and isotopic rotations are regarded as independent, the
configurational space of the proposed theory contains only one set
of group parameters and is similar to configurational space of the
MSA approach. In this sense the following construction can be
regarded as a certain modernization of the MSA approach. In order
to provide for the nondegenerate nature of the suggested change of
variables, we limit the admissible variations of the field
$\varphi(x)$ by the orthogonality conditions
\begin{equation}
<\frac{\partial}{\partial
a}[T\overline{\varphi}(Tx)]g(T\overline{\varphi}(Tx))T\delta
\varphi (Tx)>=0\label{c}\ ,
\end{equation}
where $\overline{\varphi}(x)$ is the gauge function. While
describing the collective motion, the field $\varphi(x)$ is
defined as the extremal of the Hamiltonian with fixed values of
the collective coordinates and momenta. This description can be
essentially simplified by identifying gauge
$\overline{\varphi}(x)$  with the extremal of field $\varphi(x)$.
Then the canonical momentum conjugate to the field $\varphi(x)$
vanish at the extremum point. That enables in describing purely
collective motion to put velocities $\dot\varphi(x)$ equal to zero
\cite{Ts}. In this case
\begin{equation}\label{i}
\dot{\phi}(x)=-iT\left((\hat{\boldsymbol{I}}-\hat{\boldsymbol{l}}(x))\boldsymbol{\omega}
\right)\varphi (Tx)\ ,
\end{equation}
where the velocity form $\boldsymbol{\omega}$ is defined by
expression
\begin{equation*}
T^{-1}\dot{T}=-i\hat{\boldsymbol{I}}\boldsymbol{\omega }\ .
 \label{M}
\end{equation*}
Using (\ref i), let us find the connection between canonical
momenta and velocities
\begin{equation*}
\pi_{s}\equiv \partial L
/\partial\dot{a}^{s}=-4\omega_{i}(D^{i}D^{k})\hat{\omega}_{ks},\qquad
\hat{\omega}_{ks}\equiv\partial \omega_{k}/\partial \dot{a}^{s}\ ,
\end{equation*}
where kinematic operators $\hat D^{i}_{ks}$ and symbols
$(D^{i}D^{k})$ are defined by formulas
\begin{equation*}
\hat{D}_{ks}^{i}=\frac 1 {2}\left
(\hat{I}_{ks}^{i}-\hat{l}_{ks}^{i}(x) \right )\ , \qquad
(D^{i}D^{k})\equiv
<\hat{D}^{i}\varphi(Tx)g(\varphi(Tx))\hat{D}^{k}\varphi(Tx)>\ .
\end{equation*}
Besides, using (\ref b), one can find the exact formulas,
connecting canonical momenta and the integrals of motion of the
system (\ref d)
\begin{equation}
\pi_{s}\equiv \partial L
/\partial\dot{a}^{s}=(\boldsymbol{I}T-\boldsymbol{J})_{k}\,\hat{\omega}_{ks}\
.\label{r}
\end{equation}
Thus, we obtain the functional
\begin{equation*}
H_{c}(\varphi(Tx),\sigma)\equiv \omega \frac{\partial L}{\partial
\omega}-L=-\frac 1{8}\sigma(DD)^{-1}\sigma+H(\varphi,0)\ , \qquad
\boldsymbol{\sigma}\equiv \boldsymbol{J}-\boldsymbol{I}T\ ,
\end{equation*}
which at the extremum point defines  the collective Hamiltonian of
the system. Extremal of the field $\varphi(x)$ should be
determined taking into account the constraint (\ref c). If one
takes into account the non-degeneracy of the matrice
$\hat{\omega}$ and formulas
\begin{equation}
T^{-1}\hat{\boldsymbol{I}}T=T\hat{\boldsymbol{I}}\ ,\qquad
\hat{\boldsymbol{l}}(Tx)=T\hat{\boldsymbol{l}}(x)\label{f}\ ,
\end{equation}
then in special gauge $\overline\varphi(x)=\varphi(x)$ the
constraint (\ref c) will assume the form
\begin{equation*}
<\hat{\boldsymbol{D}}_{T}\varphi(x)g(\varphi(x))\delta \varphi
(x)>=0\ ,\qquad \hat{\boldsymbol{D}}_{T}\equiv T^{-1}\hat{D}T\ .
\end{equation*}
Therefore, having created the projection operator
\begin{equation*}
\begin{split}
&\mathcal{P}^{ik}(x,x^{\prime})=-g(\varphi(x))\hat{D}^{i}_{T}
\varphi(x)>G^{-1}_{ik}<\hat{D}^{k}_{T}\varphi(x^{\prime})\ ,\\
&G^{ik}\equiv
-<\hat{D}^{i}_{T}\varphi(x)g(\varphi(x))\hat{D}^{k}_{T}\varphi(x)>\
,
\end{split}
\end{equation*}
we will find the following equation for the extremals of the field
$\varphi(x)$
\begin{equation}
(1-\mathcal{P})\frac {\delta H_{c}(\varphi (Tx),\sigma)}{\delta
\varphi}\label{e}\ .
\end{equation}
Part of this equation, containing projection operator
$\mathcal{P}$, reflects the availability of non-compensated
dynamic stresses in the system. If these stresses vanish, the
equation (\ref e) turns into equation for stationary states of the
Hamiltonian $H_{c}(\varphi (Tx),\sigma)$. Having determined the
time dependence of the matrix $T$ under the action of the
Hamiltonian $H_{c}$, one can obtain the exact dynamic solutions of
the motion equation for the field $\varphi(x)$. Let us analyze the
conditions under which the dynamic stresses vanish. Using formulas
(\ref f), one can find the obvious properties of the Hamiltonian
$H_{c}$
\begin{equation*}
H_{c}(\varphi (Tx),\sigma)=H_{c}(T\varphi (x),T\sigma).
\end{equation*}
This relationship is valid for arbitrary $\varphi(x)$, not limited
by any additional conditions. Thus, differentiating it with
respect to the parameters $a^{i}$, one can find the identity
\begin{equation}
\frac {\delta H_{c}}{\delta
\varphi(x)}\left(T^{-1}\boldsymbol{\hat l}(x)+\boldsymbol{\hat
I}\right)\varphi (x)+\frac{\partial H_{c}}{\partial
\sigma}\boldsymbol{\hat I}\sigma=0\label{g}.
\end{equation}
Next, with the help of canonical brackets
$\{a^{i},\pi_{k}\}=\delta^{i}_{k}$ , one can find the following
Poisson brackets
\begin{equation*}
\{\sigma_{i},\sigma_{k}\} =\varepsilon_{ikl}\sigma_{l}\ ,\quad
\{\sigma_{k},T_{sn}\}=\varepsilon_{ikn}T_{si}\ ,\quad
\{\sigma_{k},(T\sigma)_{i}\}=0.
\end{equation*}
Hence we derive the rates of change of the variables
$\boldsymbol\sigma, T$ under the action of the collective
Hamiltonian $H_{c}$
\begin{equation*}
\dot\sigma_{i}\equiv\{\sigma_{i},H_{c}\}=-i\frac{\delta
H_{c}}{\delta\varphi(x)}\hat{I}^{i}\varphi(x)\ ,\quad \dot
T\equiv\{T,H_{c}\}=-i\frac{\partial H_{c}}{\partial
\boldsymbol{\sigma}}T\boldsymbol{\hat I}.
\end{equation*}
Using these formulas and identity (\ref g), we obtain
\begin{equation*}
\boldsymbol{D}_{T}\varphi(x)\frac{\delta
H_{c}}{\delta\varphi(x)}=iT\boldsymbol{\dot\sigma}+\frac{i}{2}\dot{T}\boldsymbol{\sigma}.
\end{equation*}
Since spin $\boldsymbol J$ and isospin $\boldsymbol I$ are the
integrals of motion, the equation follows
\begin{equation*}
\boldsymbol{D}_{T}\varphi(x)\frac{\delta
H_{c}}{\delta\varphi(x)}=\frac{i}{2}\dot{T}(\boldsymbol{J}+\boldsymbol{I}T),
\end{equation*}
which defines dynamic stresses in the system. These stresses
vanish, if
\begin{equation}
\dot{T}(\boldsymbol{J}+\boldsymbol{I}T)=0\label{h}.
\end{equation}
We will seek a solution for the equation (\ref e) in the class of
axially-symmetrical configurations
$\boldsymbol{k}(\boldsymbol{\hat I}+\boldsymbol{\hat
J})\varphi(x)=0$, where $\boldsymbol{k}$ is unit vector in the
direction of vector $\boldsymbol{\sigma}$. In this case with the
help of (\ref b), it is easy to show that
$\boldsymbol{J}^{2}=\boldsymbol{I}^{2}$. Since the solution for
the matrix $T$ can be presented as $T=exp(i\alpha\boldsymbol{\hat
I}\boldsymbol{k}t)$, then the equation (\ref h) implies, that
dynamic functional $\boldsymbol{J}+\boldsymbol{I}T$ is a zero-mode
of the operator $\boldsymbol{\hat I}\boldsymbol{k}$. That is,
$\boldsymbol{J}+\boldsymbol{I}T=c\boldsymbol{\sigma}$. Multiplying
this equation by $\boldsymbol{\sigma}$ and taking into account
that $\boldsymbol{J}^{2}=\boldsymbol{I}^{2}$, we will find that
$c=0$, i.e., the dynamic stresses vanish on the states on which
spin and isospin are rigidly bound by the condition
\begin{equation}
\boldsymbol{J}+\boldsymbol{I}T=0.\label{p}
\end{equation}
Thus, in stationary case $\dot\varphi(x)=0$, the equation for the
field $\varphi(x)$ takes the form
\begin{equation*}
\frac{\delta H_{c}}{\delta\varphi(x)}=0\ ,\quad
H_{c}=-\frac{1}{2}J(DD)^{-1}J+H(\varphi,0)\ .
\end{equation*}
If one presents symbols $(D^{i}D^{k})$ as $(DD)=-a(1-\hat k \hat
k)-b\hat k \hat k+c\boldsymbol{I}\boldsymbol{k}$, then the inverse
matrix can be written as $(DD)^{-1}=\lambda(1-\hat k \hat
k)-b^{-1}\hat k \hat k+\gamma\boldsymbol{I}\boldsymbol{k}$.
Therefore, the collective Hamiltonian, describing synchronous
rotation of the soliton in isotopic and spin spaces, will take the
form
\begin{equation}
H_{c}=\frac{\boldsymbol{J}^{2}}{2b}+H(\varphi,0)\ ,\quad
b(\varphi)\equiv
k_{i}k_{r}(D^{i}D^{r})=-<\boldsymbol{k}\boldsymbol{\hat
I}\varphi^{s}g_{st}(\varphi)\boldsymbol{k}\boldsymbol{\hat
I}\varphi^{t}>.\label{s}
\end{equation}
The obtained Hamiltonian properly coincides with the Hamiltonian,
describing the nucleon and delta-isobar in the CDD formalism,
which takes into account an independence of the spin and isospin
rotation \cite{Ts}. At the same time, Poisson brackets algebra of
the dynamic functionals $\boldsymbol J$ in (\ref s) and algebra of
the functionals $\boldsymbol R$, defining similar Hamiltonian in
the CDD formalism, are different. It results in different quantum
description of the nucleon and delta-isobar in the considered
approach and in the scheme of the work \cite{Ts}. From the
formulas (\ref r), (\ref p) it follows that
\begin{equation*}
\boldsymbol{J}=-\frac{1}{2}\boldsymbol{\pi}\hat\omega^{-1}\ ,\quad
\boldsymbol{I}=\frac{1}{2}\boldsymbol{\pi}\hat\omega^{-1}T^{-1}\ .
\end{equation*}
These expressions for spin and isospin coincide with the
respective formulas of the work \cite{ANW} with accuracy to
parametrization. Besides, Hamiltonian  $H_{c}$ is the function of
 $\boldsymbol{J}^{2}=\boldsymbol{I}^{2}$. Therefore, quantization of the collective motion leads to the
same vectors of states for nucleon and delta-isobar as it does in
case of using MSA \cite{ANW}.

\section{Conclusion}
As a result of above described analysis, the consecutive
theoretical scheme was created whose configurational space is
similar to that of the MSA approach. The scheme reproduces the
results of the MSA approach and at the same time allows to
overcome some difficulties inherent in mentioned approach. In
particular, the method allows to take into account rotational
deformation of the nucleon and delta-isobar, doesn't require an
obligatory existence of the static soliton solution and makes
possible to find dynamic soliton states in SU(2) sigma-model. In
some sense one may say that the suggested scheme is a
modernization of the MSA approach, based upon use of the fact of
synchronous rotation of spin and isospin for the nucleon and
delta-isobar states. At the same time the given scheme can not be
regarded as an acceptable alternative to nucleon and delta-isobar
description in the CDD formalism \cite{Ts}. First of all field
fluctuations in this approach have no clear physical meaning.
These fluctuations are the mixture of true field fluctuations and
the localized time-dependent states corresponding to resonances.
In case of the CDD formalism, from the very beginning resonances
are attributed to the soliton spectrum of the system and field
fluctuations have direct sense of the low amplitude $\pi$ meson
field. For this reason the new approach gives inadequate quantum
description of the studied particles. In particular the wave
functions of these particles in MSA approach differ radically from
those in the CDD formalism.

\vspace{0,5 cm}

\end{document}